# Simulating Cognition with Quantum Computers

Hongbin Wang, Jack W. Smith, Yanlong Sun
{hwang, jwsmith, ysun}@medicine.tamhsc.edu
Biomedical Informatics, College of Medicine, Texas A&M University
2121 W. Holcombe Blvd, Houston, TX 77030 USA

**Abstract**

There are inherent limits in classical computation for it to serve as an adequate model of human cognition. In particular, non-commutativity, while ubiquitous in physics and psychology, cannot be sufficiently dealt with. We propose that we need a new mathematics that is capable of expressing more complex mathematical structures in order to tackle those hard X-problems in cognitive science. A quantum approach is advocated and we propose a way in which quantum computation might be realized in the brain.

## Introduction

In 1981, Richard Feynman delivered a keynote speech in the first conference on "physics and computation" held at MIT. In the speech, entitled "simulating physics with computers", Feynman asked if physics can be simulated by a classical universal computer. Given the fact that at that time computers have already been extensively used for physical simulation and computation at that time the question might sound strange. However, equipped with some profound new understanding regarding Turing computation and quantum mechanics, Feynman answered the question with a clear "no". He concluded, "Nature isn't classical, dammit, and if you want to make a simulation of Nature, you'd better make it quantum mechanical, and by golly it's a wonderful problem, because it doesn't look so easy." (Feynman, 1982, p.486).

We face a similar problem in cognitive science today. The tenet of cognition as computation directly catalyzed the cognitive revolution in the 1950s and has since become one of the pillars of cognitive science (Miller, 2003). Undoubtedly, an impressive body of new results have been obtained, including the magic number seven ± two (Miller, 1956) and the emergence of various executable cognitive architectures (e.g., Newell, 1990). On the other hand, unfortunately, these advances are in stark contrast with the lack of progress in answering some of the long-standing tough questions in cognitive science. Theoretically, for example, what are the implications of Gödel's incompleteness theorem on seeking a computational theory of cognition (e.g., Penrose, 1989)? Empirically, some of the fundamental psychological phenomena, including consciousness, intuition, and various so-called "cognitive biases" continue to defy a satisfactory computational description. Such theoretical and empirical dilemmas hint at "the unreasonable ineffectiveness of mathematics in cognitive science" (Poli, 1999) and partially explains the finding that in recent years cognitive science has been more and more dominated by empirical psychology (and less computational explorations) (Gentner, 2010). In a sense, we are forced to re-examine the relationship between cognition and computation and answer a similar question asked by Feynman in 1981: Can we simulate cognition by a computer? If so, what kind of computer are we going to use?

The issue is also of particular importance in the light of the recent resurgence of AI. Unprecedented capacities in various areas including machine vision, natural language processing, and game playing have been recently gained thanks to the advances in deep learning methods. Can we realize human-level intelligence by deep mining big data using deep learning with more powerful computers? Can deep learning technology finally lead to intuition and consciousness in artificial systems?

Here we intend to provide some general arguments towards answering these questions. In particular, we argue that human cognition involves more complex mathematical structures (e.g., non-commutativity) that are fundamentally beyond the expressive power of classical Turing computers. Consequentially, in order to develop a simultaneously normative and descriptive theory of cognition one has to go beyond classical set-theoretical computation. We demonstrate how formalisms in quantum theory affords structures necessary for us to model interesting cognitive phenomena. Finally, we address a common criticism of the quantum brain approach and suggest a way where quantum behavior might be realized in a warm, wet, and noisy brain. We conclude by advocating that we need to go beyond classical computers for cognitive modeling in order to gain brand new insights about how the brain and the mind work.

## The Inadequacy of Classical Computation for Cognition

Although the limits of the classical theory of computation have been well-known since its inception, computers can do so much nowadays that it's easy to forget that they were invented for what they could not do. In his 1937 paper, "On computable numbers, with an application to the Entscheidungs problem," Alan Turing defined the notion of a universal digital computer (a Turing machine), and his goal was to show that there were tasks that even the most powerful computing machine could not perform (for example, the halting problem). These problems are simply beyond computation, in accordance with the now-famous Turing-Church thesis.

A similar result was obtained at about the same time by Kurt Gödel. In 1931, through an ingenious device known as Gödel numbering, Gödel found a way to assign natural numbers in a unique way to the statements of arithmetic themselves, effectively turning numbers into statements that talk about numbers. This permitted him to prove an incompleteness theorem, which basically says that there are true statements of mathematics (theorems) which we can never formally *know* to be true.

It is interesting that although both Turing and Gödel proved that the complete body of human knowledge cannot be acquired by formal computation alone given the method's inherent limits (see Figure 1), they appear to offer different reasons for why the human mind is able to achieve the feat. According to Turing, it is unfair to compare a Turing machine and a human mind – the former runs algorithmically and never makes mistakes and the latter does "trial-and-error" and makes wild guesses all the time. "If a machine is expected to be infallible, it cannot also be intelligent". And a machine can become intelligent and human-like only if it makes no pretense at infallibility. Gödel, on the other hand, did not want to give up on the consistency of human knowledge. He suggested that "it remains possible that there may exist (and even be empirically discovered) a theorem-proving machine which in fact is equivalent to mathematical intuition, but cannot be proved to be so, nor can be proved to yield only correct theorems of finitary number theory".

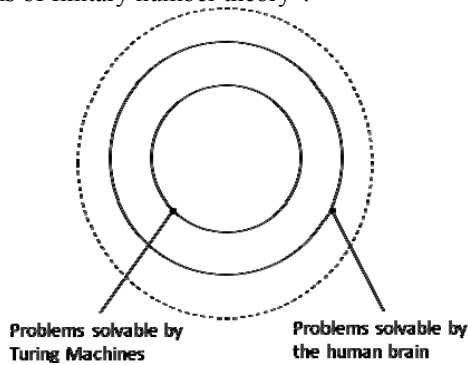

Figure 1. A hypothetical problem complexity space.

In the end of his now famous 1960 article entitled "The unreasonable effectiveness of mathematics in the natural sciences", Eugene Wigner wondered "if we could, some day, establish a [mathematical] theory of the phenomena of consciousness, or of biology, which would be as coherent and convincing as our present theories of the inanimate world". Half a century later, Wigner's hope apparently hasn't been fulfilled. The list of "cognitive biases" on Wikipedia is getting longer and longer. Human "irrationality" seems everywhere. Of course irrationality in all these cases is defined as human deviation from classical logic and standard mathematics. Decades of research in psychology appears to show beyond doubt that such deviation is systematic and abundant. The discrepancy shown in Figure 1 hints at an explanation: we might have used a mathematics that is not powerful enough to capture all the complexity of human cognition.

There have been efforts to extend the classical theory of computation – its history is almost as long as the classical computational theory itself. Various models of computation that can compute functions not effectively computable in the Church-Turing thesis sense, often called hypercomputation or super-Turing computation, have been suggested (see https://en.wikipedia.org/wiki/Hypercomputation).

One notable hypercomputational model is the Blum-Shub-Smale (BSS) machine (Blum et al., 1997), also called a real computer. As we know, a classical computer depends on discrete symbols (e.g., 0s and 1s) to encode information and presupposes that all the underlying sets are countable (one-to-one correspondence to natural numbers N). A real computer is able to handle real numbers (R, a continuum) and therefore can answer questions about subsets which are uncountable (e.g., "is the Mandelbrot set decidable?"). While it has been shown that real computation can be directly applied to problems in numerical analysis and scientific computing, it is not clear if it helps reduce the discrepancy shown in Figure 1 in any fundamental way. We argue that the extension from N to R, as significant as it may seem, remains inadequate to handle some of the toughest problems (see below) in cognitive science.

## X-problems and Non-Commutative Observables

Following Penrose's practice in physics (Penrose, 1997), it is helpful to distinguish two classes of problems in cognitive science that long for answers. One class can be called Z-problems (for puZZle), which refer to those empirical findings that are puzzling but somewhat explainable in classical computational terms. Examples of Z-problem include the distinction of short-term memory and long-term memory, the concept of working memory capacity, skill acquisition by forming and tuning if-then production rules, and attention through bottom-up and top-down controls. Another class of problems can be called X-problems (for paradoXes), referring to those empirical findings that are so mysterious that they seem to defy classical mathematical descriptions. Examples of this class include consciousness and awareness, intuition, feeling, gestalt phenomena in visual perception, and various so-called "cognitive biases" in human judgment and decision-making, to name a few. Discrediting them as ephemeral and unworthy, or simply labeling them as "human biases and heuristics", or suggesting ad-hoc patched explanations, is inadequate. These phenomena are functions of the human brain/body, which resulted from millions of years of evolution and adaption. They deserve more rigorous and more systematic treatments and it is fair to say that cognitive science so far has fallen short in this regard.

Tackling X-problems requires us to clearly recognize the fundamental limits of the currently dominating classical theory of computation as a model of human cognition and seek a new mathematics that can describe human behavior rather than treat human behavior as exceptions or deviations.

What the new mathematics will look like remains to explored. One necessary condition seems to be that it has to be equipped with some extra non-trivial hidden mathematical structures that can afford complex emerging phenomena. In this paper we are advocating a particular aspect of new mathematics that we deem essential and even magical to tackle some of the X-problems in cognitive science. This aspect has to do with non-commuting observables. Briefly, non-commutativity (e.g., ab ≠ ba) is ubiquitous in physics and psychology, but the classical theories of computation, operating on the 1-dimensional real space (N in Turing computer and R in real computation), leaves little if any room for non-commutativity.

Generally speaking, commutativity concerns the order effect of an operation. We may tend to think that commutativity is a normalcy (for example, we learn in schools that number multiplication is commutative, 2x3=3x2) but not all operations in nature are commutative. In fact, in mathematics commutativity is typically just a nice feature that needs to be specially denoted. For example, in a canonical group, a mathematical structure where a multiplication is defined among a set of elements, commutativity is not required. Those special groups where multiplication is commutative are then called abelian groups. In mathematics, a commutator can be defined to indicate the extent to which a certain binary operation fails to be commutative.

Non-commutativity is abundant in physics. As a matter of fact, realizing that observables can be non-commuting and developing mathematical formalism that can handle non-commutative variables have been some of the most important achievements that directly catalyzed the birth of quantum physics about a century ago. Given the remarkable similarity to the situation we face in cognitive science today, let us elaborate a little.

We know that mechanics concerns the movement of objects in space and time. Two variables are used to describe the state of a moving object, the position (x, assuming a 1-dimensional space here) and the velocity (v). In the Hamiltonian formalization of classical mechanics, velocity is replaced by momentum p (=mv). Therefore the state of the system can be completely represented by a point in so-called phase space, which in this case is just a 2-dimensional (x and p) manifold. And the dynamics is governed by a smooth Hamiltonian function, H(p,x), defined on the manifold.

This classical picture of mechanics basically treats x and p as if they are two independent variables, and each can be independently measured. One of the key insights that fueled the mathematical basis of the quantum revolution is the realization that x and p are not independent. In fact, in quantum mechanics p is identified with a differential operator with respect to the position. With details omitted, the consequence of doing so is that x and p are non-commuting variables, and their relationship can be summarized as a remarkable and non-zero commutator:

$$[x, p] = i\hbar \mathbf{1}$$

where $i$ is the imaginary unit, $\hbar$ is the reduced Planck constant. We cannot over-emphasize the importance of this relation. This relation leads (or is equivalent) to many essential principles of quantum physics, such as the Heisenberg uncertainty principle, and directly gives rise to a range of quantum phenomena (e.g., the wave-particle duality).

The non-commuting relation between x and p is no accident. They reflect some of the fundamental conjugating relations in nature, such as movement vs stillness, time and energy, global vs local, differentiation vs integration. It is only remarkable that they are linked by such fundamental constants as $i$ and $\hbar$. Together they depict a multi-dimensional non-flat geometry of the physical world.

It would have been unfortunate if psychology were not full of such non-commuting observables. Mental objects can only be more complex and possibly live in a higher-dimensional spacetime. In a psychological experiment (as well as in everyday experience), we measure many variables – judgment on question 1, decision on question 2, reaction time for trial x, accuracy for person y, feeling on day z, etc. Then we do statistics on these variables. We often treat them as if they live on a huge Cartesian product space and each dimension is independent of another (which is multi-dimensional in appearance, but can be projected to 1-dimensional N via Gödel numbering). Doing so greatly reduces dimensionality and conceals/distorts the underlying structures. In cases where the non-commuting nature of variables somehow reveals itself in statistical results one way or another, we then dismiss them as cognitive biases, intuition, satisficing, or human irrationality.

A new mathematics of human cognition therefore calls for that we go beyond the symbolic set-theoretical classical theory of computation, and treat non-commuting observables in psychology seriously in order to truly describe rather than prescribe human cognition. Fortunately, we probably don't have to develop a brand-new mathematics and instead we can start with some remarkable techniques mathematicians and physicists developed in the past two centuries or so that can handle non-flat spaces of various dimensions. A quantum information theoretical approach to cognition is a worthy choice.

## Towards a Quantum Mind

Attempting to understand the mind from a quantum theoretical perspective is not new. Wolfgang Pauli's joint adventure with Carl Jung in 1930s in studying synchronicity and telepathy has been well known. Erwin Schrodinger, a founder of quantum physics, published "what is life" in 1944 and delivered the Tarner Lectures on "Mind and Matter" in Cambridge in 1954. David Bohm and Rogers Penrose, both prominent physicists, pioneered more modern efforts to understand what consciousness is and how it can arise in the brain from a quantum-theoretical perspective

(Bohm, 1980; Penrose, 1989). More recently, Jerome Busemeyer and colleagues have been promoting a quantum treatment to cognition (Busemeyer and Bruza, 2012), and Matthew Fisher has proposed a quantum brain project (Fisher, 2015).

Our new math is so motivated by and consistent with quantum physics and quantum information theory (Nielsen and Chuang, 2000) that we certainly do not mind to align ourselves with this distinguished group of researchers in advocating a general quantum approach to the brain and the mind. However, we would like to make the following general distinction.

First, our goal is to seek a new mathematical theory of the human mind that is simultaneously normative (self-consistent, systematic, rather than patched), descriptive (descriptive rather than prescriptive of human behavior), and biologically realistic (neuronally implementable). Our theory presupposes a high-dimensional non-flat mental space that cannot be reduced to lower dimensions without losing information. This is consistent with the quantum formalism of physics. We take quantum theory seriously but not literally since we are dealing with mental space, which is likely quite different from physical space.

Second, we are serious in exploring how our quantum computation theory can be realized biologically in the human brain. It is in this aspect we are different from the more behavior-focused quantum cognition approach. For example, Busemeyer clearly stated, "We are not interested in physics; neither do we claim the brain is a quantum computer. Our interest lies solely in the application of mathematical principles from quantum probability theory to behavioral data observed in social and behavioral sciences".

Third, while we claim that the brain is a quantum computer, we do not propose at this stage how it is literally realized as a quantum level of phenomena. We are aware of those rather heated debates regarding this issue. Our position is that we do not need, at least until we know more about neuroscience, to identify an atomic or molecular level of mechanisms for quantum realization in the brain. Instead, we suggest that because the brain has so many neurons, each of which computes in its own frame of reference, quantum like behavior emerges naturally when all sub-systems have to be integrated at a higher level. We will briefly discuss the issue later.

In the rest of the paper, we will provide a case study to demonstrate how tossing a fair coin, probably the simplest kind of probabilistic event, can be represented in a quantum mind, and why the gambler's fallacy is a quantum like behavior, and how it could emerge in the human brain.

## Coin Tossing in a Quantum Mind

Tossing a coin, which could land on head (H) or tail (T), is modeled by a bit in the classical theory. The whole phase space consists of two points (on R), 0 (for H) and 1 (for T). Tossing a fair coin can be represented by a state with binary choices, H and T, each with a ½ probability. Formally,

$$S = \tfrac{1}{2} H + \tfrac{1}{2} T$$

In quantum theory, a coin tossing can be represented by a Qbit (a quantum bit), which lives in a complex 2-dimensional Hilbert space (see Figure 2). Every point on the Bloch sphere is a possible state of a Qbit. A Qbit is similar to a classic bit in that when measured it reveals itself as 0 or 1. However, roughly speaking, a Qbit contains much more (quantum) information than a classic bit – every point (there are infinite number of them) on the equator unit circle of the Bloch sphere represents a possible state of fair coin tossing. So tossing a fair coin can be represented equally well by the following two states (among infinitely others).

$$|\psi 1\rangle = \tfrac{1}{\sqrt{2}} \times |H\rangle + \tfrac{1}{\sqrt{2}} \times |T\rangle,$$
$$|\psi 2\rangle = \tfrac{1}{\sqrt{2}} i \times |H\rangle + \tfrac{1}{\sqrt{2}} i \times |T\rangle$$

These representations are called wavefunctions. The coefficients are complex numbers, and each modulus squared represents a corresponding probability.

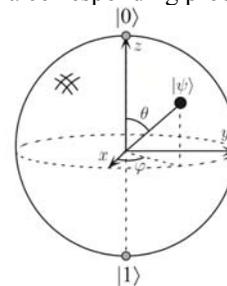

Figure 2. Bloch sphere representation of a quantum bit.

This representation certainly affords the expression of more degrees of freedom. These additional degrees of freedom seem to represent phase information, which is supposed to live on a different dimension. The resulting states are traditionally translated to something like, "the coin is at a superposition state of head and tail at the same time". This kind of description causes a lot of confusion for non-quantum physicists. It seems that our natural language lacks the necessary expressive power to even describe these intimately confounded states, let alone to distinguish them. The questions are: Can our brain do that? And what is the possible psychological relevance of the phase information, if any?

Any sane psychologist would probably quickly reject this nonsense: "why do I need an infinite number of possible states to represent a simple half-and-half coin tossing, which I even do not know how to tell them apart?" This is certainly a fair question, and here are some counter-arguments. First of all, the brain has so many neurons, and we do not know, based on our current knowledge of neuroscience, how each and every one of them is really representing for a given perception. It is certainly possible that the phase information is represented somehow by neurons or neuron groups. Second, unlike experimental quantum physicists who design sophisticated and expensive apparatus that allow them to do various precise and fine-grained measurement on quantum particles in their labs, psychologists are far inferior in their capacity to measure neuronal or mental states. Tools in their arsenal, including introspection, verbal report, reaction time, or even neuroimaging (fMRI or EEG), often lead to rough, coarse, imprecise, and indirect measures. Despite this deficit, it is still possible that this extra hidden phase information may manage to reveal itself in some ways that can be detected empirically, if we look carefully. We would like to argue that the Gambler's fallacy is such a manifestation.

## Gambler's Fallacy is a Quantum Phenomenon

Suppose you see an H when you toss a fair coin. Which side do you think the coin will land on in your next toss? Classical probability theory will predict a 50-50 chance for H and T (as it is apparently an independent Bernoulli process). But a large body of empirical research has robustly shown that people tend to think that T is more likely ($p \approx 0.6$), committing a so-called Gambler's Fallacy. In general, it refers to the belief that chance is a self-correcting process where a random event is more likely to occur because it has not happened for a period of time. For decades, this fallacy has been regarded as a prime example of human irrationality and thought to have originated from a cognitive bias called the "representativeness heuristic" (Tversky and Kahneman, 1974). Efforts to patch classical probability theory to explain the phenomenon have been unsatisfactory (e.g., Hahn and Warren, 2009; also see Sun, Tweney, and Wang, 2010).

But chance is not the only observable for an uncertain state of an object. It is also possible to measure the state in time. The difference between chance (measuring how likely an event is to occur) and time (measuring when an event is to occur) can be better revealed by examining this question, is a more likely event going to happen sooner? Common reasoning may lead to a positive answer. It turns out, however, that chance and time are really two different measures for an uncertain event – they are not always correlated. In particular, it can be shown that while for a fair coin HH (two heads in a row) and HT (a head followed by a tail) are equally likely ($p=1/4$), their waiting time is different: on average, it takes 4 tosses to see the first HT but it takes 6 tosses to see the first HH. Somehow HH is harder to come by than HT even though they have the same chance (Sun & Wang, 2010a; Sun & Wang, 2010b).

More critically, it seems that chance and waiting time are two non-commuting variables. While both variables concern the temporal distance between the consecutive re-occurrence of an uncertain event, the former measures the mean distance and the latter measures its variance. Variance can be regarded as a phase factor. For a classical coin, the non-commutativity of the two variables may not induce any observable effects for a single toss. When two consecutive tosses are examined together, phase factors manifest an observable difference in terms of waiting time.

One may wonder why we have to examine two consecutive tosses together when we were taught (and we believed) that the two tosses are completely independent and identical Bernoulli trials. Where does the emerging effect of connection come from? If we say that the seemingly mysterious connection between the two consecutive and classically perfectly independent coin tosses comes from quantum entanglement, people may be suspicious. But according to quantum physics, there are actually many more quantum entangled states than quantum un-entangled states. In this case, it can be shown that simply pondering in the mind if the two independent tosses will be landing on the same (or different) side is enough to entangle the two tosses (Sun et al., 2018). Formally, let us then use

$$|+\rangle = \frac{1}{\sqrt{2}}|H\rangle + \frac{1}{\sqrt{2}}|T\rangle$$

to represent a single fair coin toss. A pair of two consecutive independent tosses can be represented as a product state that contains no correlation,

$$|\psi\rangle = |+\rangle |+\rangle$$

For neurons to represent the state that the two tosses will land on the same side (HH or TT), the measurement can be represented as a projector:

$$\Pi^{same} = |HH\rangle\langle HH| + |TT\rangle\langle TT|$$

It is remarkable that when we apply the projector to $|\psi\rangle$ we obtain a maximally entangled state:

$$\frac{\Pi^{same}|\psi\rangle}{\sqrt{\langle\psi|\Pi^{same}|\psi\rangle}} = \frac{1}{\sqrt{2}}(|HH\rangle + |TT\rangle) = |\beta_{00}\rangle$$

This exercise demonstrates that quantum entanglement is not hard to come by in the mind if we treat various hypothesis generation processes as intermediate measurements. And our mind is doing hypothesis generation all the time, a process called abduction. Under certain conditions, originally hidden information starts to reveal itself through an observable effect, leading to the Gambler's fallacy.

## Gambler's Fallacy in the Brain

Question remains regarding if and how such quantum operations can be realized in the human brain. Quantum computation with true quantum level devices seems to require strictly isolated physical environments which the brain lacks. Several proposals exist. A well-known one is the Orch-OR theory proposed by Penrose and Hameroff (see Hameroff and Penrose, 2014 for a recent review). According to this theory, quantum computation takes place in microtubules in neurons and periodically orchestrated objective reduction occurs due to quantum gravity which gives rise to consciousness. The theory is seriously challenged by Max Tegmark, a MIT physicist, whose calculation shows that in the brain's hot, wet and noisy environment quantum coherence won't last longer than $10^{-20}$ of a second, too short for the proposed mechanism to work (Tegmark, 1999). The debate continues as of today.

More recently, Fisher (2015) explored the possibility of quantum processing in the brain with nuclear spins. In particular, phosphorus is identified as the unique biological element with a nuclear spin that can serve as a neural qubit. Using the Posner molecule, Fisher shows that the neural qubits entanglement can be protected for much longer time

(e.g., minutes and hours) and thereby can serve as a (working) quantum-memory.

Remarkably, according to the idea we suggested above, it seems that we don't have to go to atomic and molecular levels to see if the brain is a quantum computer. It is possible that we can observe traces of quantum computation at the neuron or neural networks levels.

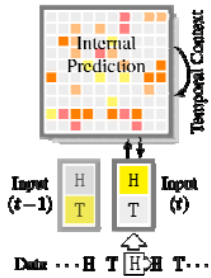

Figure 3. A biologically realistic neural network model reported in Sun et al. (2015). Initially undifferentiated hidden neurons started to differentiate after experiencing sequences of binary events.

One of our early works seems to support this hypothesis (Sun et al., 2015). Briefly, we developed a simple biologically-realistic recurrent neural network, and trained it, via unsupervised Hebbian learning, on binary sequences of random coin tossing (see Figure 3). We found that with training initially undifferentiated hidden neurons started to differentiate – some became H detectors and some became T detectors. Critically, through an activation-based receptive field analysis, we also found neurons that seemed to be sensitive to consecutive tosses. In particular, some neurons were more sensitive to repetition (R) (HH or TT), and some neurons were more sensitive to alternation (A) (HT or TH). Somehow it appears that the brain, through overlapping neuron groups, had automatically entangled consecutive inputs. Most remarkably, when we counted the number of R detectors and the number of A detectors in the hidden layer, we found an R/A ratio of 0.7, indicating that there were more A detectors than H detectors. This is "paradoxical" since the chance of R and A were exactly the same in the input sequences! However, we have shown that the ratio was perfectly correlated with the waiting time statistics of respective patterns of tosses for a fair coin, suggesting that the neural network learned to be prone to the Gambler's fallacy just as humans are.

## Conclusions

In this paper we advocate that a new mathematics is needed to tackle X-problems in cognitive science. A quantum mind approach is explored. We show that the quantum formalism affords the capacity to represent mathematical structures that are more complex than those permitted by the classical theory of computation. In particular, we show that non-commutativity, while ubiquitous in physics and psychology, cannot be adequately handled by classical Turing computers. We propose that the brain performs quantum computation and that we should use quantum computation to model the mind.

## Acknowledgments

This work was partially supported by an Air Force Office of Scientific Research (AFOSR) grant (FA9550-12-1-0457), and an Office of Naval Research (ONR) grant (N00014-16-1-2111).